\documentclass{elsart}

\usepackage{epsfig}

\usepackage{amssymb}
\begin{document}
\begin{frontmatter}

\title{Opinion dynamics  and  synchronization \\in a network of 
scientific collaborations}


\author[label1]{Alessandro Pluchino*},  \corauth[cor1]{Corresponding author: alessandro.pluchino@ct.infn.it}   
\author[label2]{Stefano Boccaletti}, 
\author[label1]{Vito Latora},   
\author[label1]{Andrea Rapisarda} 

\address[label1]{Dipartimento di Fisica e Astronomia,  Universit\'a di Catania,\\
and INFN sezione di Catania, Via S. Sofia 64,  I-95123 Catania, Italy}
\address[label2]{ CNR-Istituto dei Sistemi Complessi, Largo E.
Fermi, 6, 50125 Florence, Italy} 

\begin{abstract}


In this paper we discuss opinion dynamics in the {\it Opinion Changing Rate (OCR)} model, 
recently proposed in Ref.~\cite{ocr}. 
The OCR model allows to study whether and how a group of social 
agents, with a different intrinsic tendency ($rate$) to change opinion,  
finds agreement. In particular, we implement the OCR model on a small 
graph describing the topology of a real social system. The nodes of the 
graph are  scientists partecipating to the Tepoztl\'an conference, celebrating Alberto Robledo's 60th birthday, and 
the links are based on coauthorship in scientific papers.  
We study how opinions evolve in time according to the frequency rates of 
the nodes, to the coupling term, and also to the presence of group 
structures. 
\end{abstract}

%
\end{frontmatter}

\section{Introduction}

\label{intro}

In the last years
there has been an increasing interest  in statistical physics for 
interdisciplinary applications.  Not only 
biology, economy and geology, but also  soft sciences \cite{general} 
like  sociology or cognitive sciences 
have been involved. An effort in this sense
has also been advocated in order to strengthen the scientific aspect of these disciplines \cite{nature-soft}. At the same time,  the study of complex networks, concerning both their structure and their dynamics,  
has seen an explosive evolution \cite{rev1,rev2,rev3}. 
This field  is probably only in its infancy and 
will likely reinforce the  interdisciplinary new directions of contemporary statistical physics.
Within this scenario, many {\em sociophysics} papers have been published and new models have 
been proposed  for studying in particular opinion dynamics and 
consensus formation \cite{stauff1,HK,SDG,bennaim,stauff2,vector}. 
Although in many cases such models offer at the moment an oversimplified description 
of a real social system, they can be hopefully very useful in the 
long term. 
In this paper we discuss   opinion formation mechanisms, and in particular 
we study the implementation on a real social network of a 
recently proposed  model, 
the {\it Opinion Changing Rate} (OCR)  model  \cite{ocr}.
The latter is a modified version of the Kuramoto model \cite{kuramoto_model,k1,strogatz} 
adapted to the social context. The OCR model 
allows to explore the possible role of synchronization in the process of opinion formation. 
In previous works we have considered a group of fully coupled agents with a 
different natural tendency ($rate$) to change opinion.  
Here we extend the study to a system of agents on a small graph representing a 
real social network, namely the network of scientific collaborations
among statistical physicists in which Alberto Robledo is involved. 
\\
The paper is organized as follows. In section 2 we review the main 
features of the OCR model. In section 3 we explain how the network 
has been constructed. There we study its structural properties with particular 
attention to the subdivision in community structures and we 
discuss the results of numerical simulations of the 
the OCR model on the network. 
Conclusions are drawn in section 4.

\section{The Opinion Changing Rate (OCR) model}

Many of the most popular opinion formation models
have the limitation of not
taking into account the individual inclination to change,
a realistic feature of any individual.  
In fact each one of us changes idea, habits, style of
life or way of thinking in a different way, with a different velocity.
There are conservative people
that strongly tend to maintain their opinion or their style 
of life against everything and everyone.
There are more flexible people that change ideas very easily
and follow the current fashions  and trends. Finally,  there are
those who run faster than the rest of the world anticipating
the others. These different tendencies can be interpreted 
as a continuous spectrum of different degrees of natural
inclination to changes. 
\\
In a recent paper \cite{ocr} we showed how such a personal inclination
to change, randomly   distributed in a group of individuals, 
can affect the opinion dynamics of the group itself.
Switching from  the question 
{\it "Could agents with initial different opinions reach a final  agreement?"} 
into the more realistic  one  
{\it "Could agents with a  different
 natural tendency to change opinion reach a final agreement?"},
 we introduced a new idea, the natural
opinion changing rate, that is very similar to the
characteristic frequency of an oscillator. In such a way, 
one can treat consensus as a peculiar kind of synchronization 
(frequency locking) \cite{boc02}, a phenomenon which has been very well studied in different contexts
by means of the Kuramoto model\cite{kuramoto_model,k1,strogatz}. 
\\
The Kuramoto model of coupled oscillators is one of the simplest 
and most successful models for synchronization. 
It is simple enough to be analytically solvable, still retaining 
the basic principles to produce a rich variety of dynamical regimes
and synchronization patterns. 
The most interesting feature of the model is that,
despite the difference in the natural frequencies of the
oscillators, it exhibits a
spontaneous transition from incoherence to collective synchronization
beyond  a certain threshold of the coupling strength \cite{strogatz}.
The existence of such a critical threshold for synchronization is
very similar to the consensus threshold found in the majority of the opinion formation models.
Thus we modified the Kuramoto model in order to study  synchronization mechanisms in consensus formation.
In our model each oscillator represents an agent corresponding to a node of a given network and
the topology of the network fixes the neighborhood ${\bf K}_i$ of every agent.
The dynamics of a system of $N$ individuals is governed by the following set of differential equations:
\begin{equation}
    \dot{x_i} (t)  = \omega_i + \frac{\sigma}{k_i}\sum_{j\in {\bf K}_i}
      ~\alpha \sin ( x_j  - x_i ) e^{- \alpha |x_j  - x_i| }  ~~~~~i=1,\dots,N
\label{OCR_eq1}
\end{equation}
where $x_i (t)$ is the opinion of the $i$th agent at time $t$. 
Here the opinions have a very general meaning and can represent
the style of life, the way of thinking or of dressing etc, 
thus they can be usefully represented by means of unlimited real numbers 
$x_i \in ] -\infty +\infty[ ~~ \forall i=1,...,N$.
Opinions interact by means of the coupling term, where
$\sigma$ is the coupling strength and $k_i$ is the degree 
(i.e. the number of first neighbors) of each agent. 
The exponential factor in the coupling term,
tuned by the parameter $\alpha$, 
ensures that opinions will not influence each other any longer 
when the reciprocal distance exceeds a certain threshold.
This is perhaps the most remarkable feature of the OCR model with respect to the
Kuramoto model, since it allows the system to reach an asymptotic stationary state 
where the configuration of opinions does not vary any longer.
The parameter $\alpha$ appears also as a factor of the sine in the coupling term and
simply rescales the range of the coupling strenght.
\footnote{Please notice that, due to a misprint, this factor $\alpha$ before the sine term 
is missing in formula (7) of ref.\cite{ocr}} 
We typically adopted the value $\alpha$=3, which ensures a consistent behavior
of the exponential decay.
Finally, the $\omega_i$'s - corresponding to the {\it natural frequencies} of the 
oscillators in the Kuramoto model - represent here the so-called
\textit{natural opinion changing rates} (\textit{ocr}), i.e.
the intrinsic inclinations of the agents
to change their opinions. 
For this reason we called our model: the {\it Opinion Changing Rate}
(OCR) model \cite{ocr}. 
The values $\omega_i$'s, which do not depend on time, are distributed in a uniform random
way   with an average $\omega_0$.
In this way we can simulate the behaviour of  both conservative individuals,
characterized by small values of $\omega_i$ ($< \omega_0$), and more flexible people, 
with high values of $\omega_i$ ($> \omega_0$). Agents going against the mainstream  
can be also simulated, by choosing negative values for their \textit{ocr}.

\begin{figure}
\begin{center}
\epsfig{figure=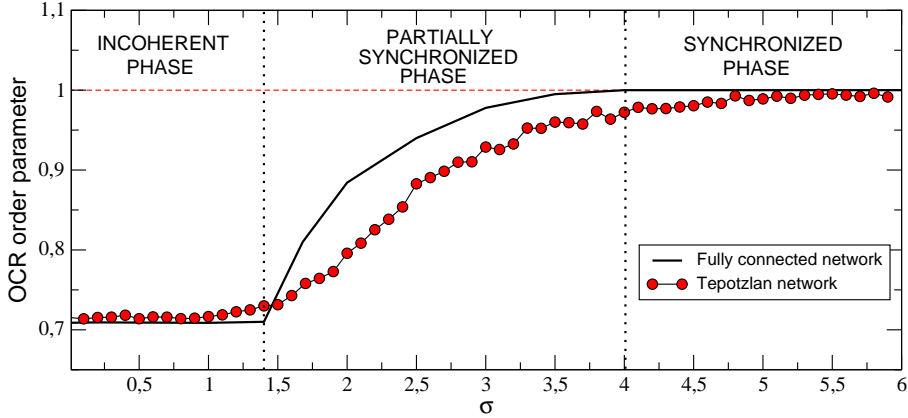,width=12.0truecm,angle=0}
\end{center}
\caption{Phase diagram of the OCR model. The order parameter is reported as 
a function of the coupling constant for a fully connected system of 
$N=1000$ agents (full line) and for the scientific collaboration 
network under investigation (full circles). In the latter case, an average over 
100 realizations was performed. See text for further details. 
}
\end{figure}

In Ref.\cite{ocr}, we studied the opinion dynamics of the OCR model considering 
a fully connected network.
The numerical simulations were performed typically
with $N=1000$ agents and with a uniform distribution of the
initial individual opinions $x_i(t=0)$ in the range [-1,1].
The natural \textit{ocr} $\omega_i$ were
taken from a uniform distribution in the range [0,1]. 
By solving numerically the set of ordinary differential equations
(\ref{OCR_eq1}) as a function of the coupling strength $\sigma$, we observed 
a transition from an incoherent phase (for $\sigma < \sigma_c$), in which people 
tend to preserve different opinions according to  their natural rate $\omega_i$, 
to a partially synchronized phase, where people share a small number of opinions, 
and, finally, to a fully synchronized one (for $\sigma >> \sigma_c$) in which all the people change 
opinion with the same rate and share a common social trend.
In order to measure the degree of synchronization of the system,
we adopted an order parameter related to the standard deviation of the
opinion changing rates and defined as 
$R(t) = 1 - \sqrt{ \frac{1}{N} \sum_{j=1}^N (\dot{x}_j(t) - \dot{X}(t))^{2}}$,
where $\dot{X}(t)$ is the average over all individuals of $\dot{x_j}(t)$.
From such a definition it follows that
$R=1$ in the fully synchronized phase and $R<1$ in the incoherent or 
partially synchronized phase.
\\
In Fig.1 we report as a full line the asymptotic (stationary) value $R_\infty$ of the order parameter 
as a function of $\sigma$ for a fully connected system, 
as reported in Ref.\cite{ocr}. A phase transition occurs at $\sigma_c \sim 1.4$.
A further analysis of the model showed that in the region $1.5 < \sigma < 2.5$ 
(belonging to the partially synchronized phase) an equilibrium between 
conservative and progressist components (a sort of  bipolarism) can be observed. 
Conversely, out of this window, the system rapidly reaches a final configuration
consisting in many small opinion clusters (anarchy) or in a single large 
cluster (complete consensus). 
Moreover, starting the system with all the agents sharing the same opinion 
and for $\sigma \sim 1.5$,  
one observes an interesting kind of metastability: even though we are in the partially
synchronized phase, the system relaxes to the partially synchronized state 
only after a metastable regime, where the opinions remain synchronized. 
The duration of such a metastable regime  
was observed to diverge when the value of $\sigma$ approaches $1.62$.

\begin{figure}
\label{network}
\begin{center}
\epsfig{figure=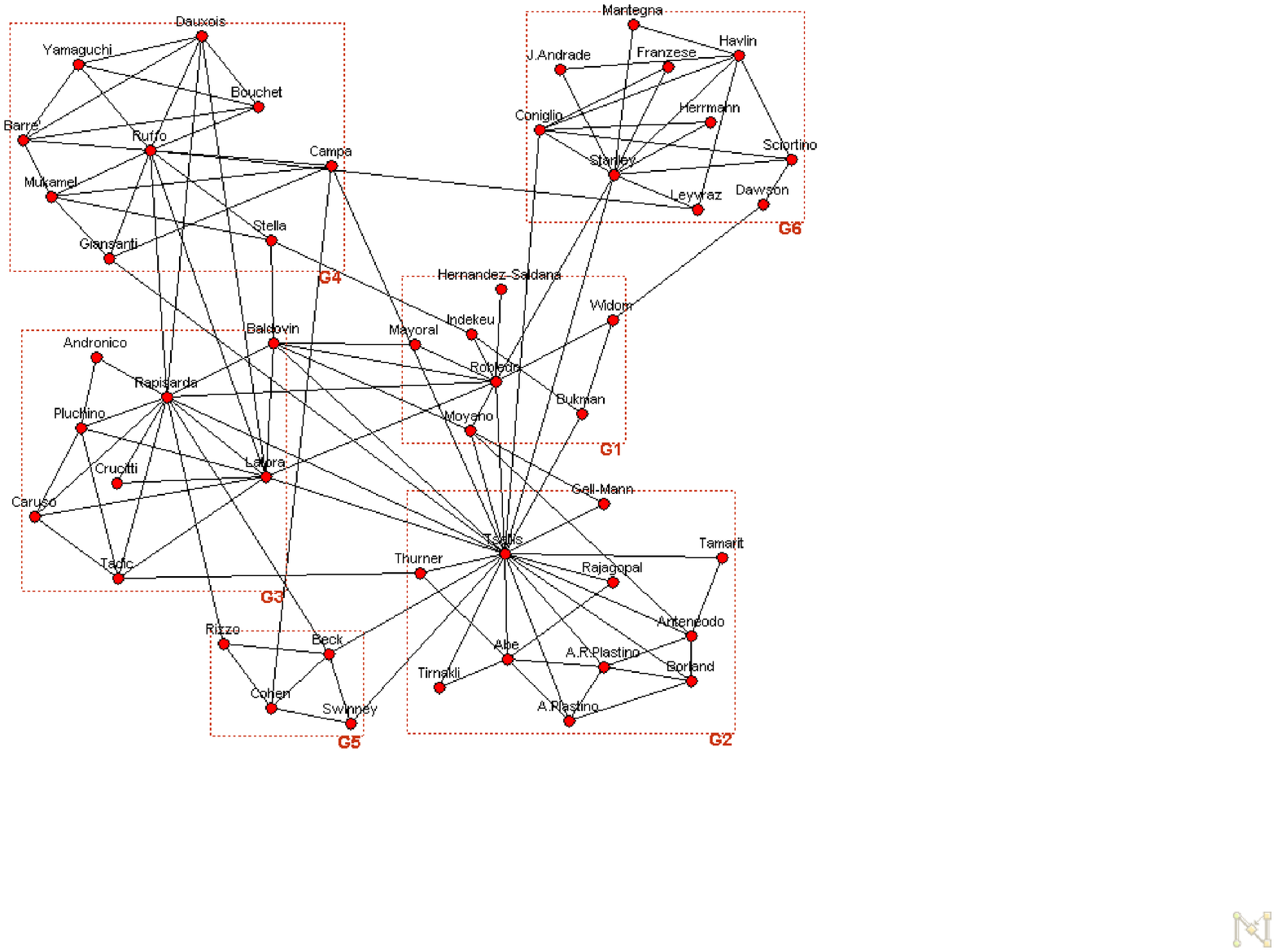,width=14.truecm,angle=-90}
\end{center}
\caption{The graph of the \textit{Tepoztl\'an Network} (TN). The 
graph under investigation is inspired to Alberto Robledo's scientific 
collaboration network. The graph is made of $N=49$ scientists and $K=117$ links 
(between couples of scientists) 
representing the existence of coauthored publications. The graph shows a clear 
division in 6 communities (groups $G_1$, $G_2$, $G_3$, $G_4$, $G_5$ and $G_6$), 
as indicated by the six squares reporting the results of 
the GN algorithm \protect\cite{GN}. See text for more details.
}
\end{figure}
\begin{table}
\begin{center}
\caption{ List of the six groups $G_1$, $G_2$, $G_3$, $G_4$, $G_5$ and $G_6$ 
of the Tepoztl\'an Network,  
as resulting from the GN algorithm for the detection of community structures \cite{GN}. 
The groups are the same as those indicated by dashed squares in Fig.~2. 
See text for details.}
\begin{tabular} {ll}
\hline 
\hline 
Group 1 (G1)  & Bukman, Hernandez-Saldana, Indekeu, Majoral, Moyano,\\
              & Robledo, Widom\\
Group 2 (G2)  & Abe, Anteneodo, Borland, Gell-Mann, A.Plastino, A.R.Plastino,\\
              & Rajagopal, Tamarit, Thurner, Tirnakli, Tsallis\\
Group 3 (G3)  & Andronico, Baldovin, Caruso, Crucitti, Latora, Pluchino,\\
              & Rapisarda, Tadic\\
Group 4 (G4)  & Barr\'e, Bouchet, Dauxois, Campa, Giansanti, Mukamel, Ruffo,\\
              & Stella, Yamaguchi\\
Group 5 (G5)  & Cohen, Beck, Rizzo, Swinney \\
Group 6 (G6)  & J.Andrade, Coniglio, Dawson, Franzese, Havlin, Herrmann,\\ 
              & Leyvraz, Mantegna, Sciortino, Stanley\\
\hline
\end{tabular} 
\end{center}
\end{table} 


\bigskip
\section{The OCR model on a real network of scientific collaboration}

In this section we study the dynamical evolution of the OCR model on the 
topology of a real social network, namely 
the network of scientific collaborations shown in Fig.2. Such a network is made of 
$N=49$ nodes and $K=117$ links. 
The nodes of the network represent some of the people 
who attended the Tepoztl\'an conference, celebrating Alberto Robledo's 60th birthday, 
including also other scientists who collaborate with them. 
For this reason we named it the \textit{Tepoztl\'an Network} (TN). 
The graph has been constructed by using information on coauthored scientific 
publications from the  cond-mat archive at arXiv.org and from the web engine {\it Google scholar}. 
The resulting graph is simple and unweighted \cite{rev3}. In fact, 
a link between node $i$ and node $j$ indicates that the two respective agents share at least 
one scientific preprint or co-editorship, which reveals the existence of 
a scientific collaboration.  
Of course some links could be likely missing, but it is not very important here
 if such a network is  fully realistic or not.  Actually what we want to do is to see how groups of collaborating people "sharing the same opinion" or the same 
scientific interests evolve in time according to the natural tendency
to change opinion (i.e. the natural \textit{ocr}) of their members, the topology of the network and 
the strength of interaction $\sigma$.
\\
In order to find the best modular division of the nodes of the graph 
into groups or communities (i.e. subsets of nodes which are more densely linked when 
compared to the rest of the network),   
we have used the Girvan-Newman (GN) algorithm for the detection 
of community structures \cite{GN}.
This is a hierarchical divisive algorithm, based on the progressive removal 
of the edges with the highest score of \textit{betweenness}, the latter being a measure 
of the fraction of shortest paths that are making use of a given edge.
The GN algorithm produces a hierarchical tree of communities starting with a single community 
including all the nodes of the graph, and ending into a set of $N$ communities 
of isolated nodes. 
But which of these nested subdivisions describes the real community structure 
of the network ? 
\\
To answer this question it was introduced the so-called
\textit{modularity} $Q$ \cite{GN}, a variable that quantifies the 
degree of correlation between the probability of having an edge joining
two sites and the fact that the sites belong to the same community.
Actually, given an arbitrary network
and an arbitrary partition of that network into $n$ communities, it
is possible to define a $n \times n$ size matrix ${\mathbf
e}$ whose elements $e_{ij}$ represent the fraction of total links
starting at a node in partition $i$ and ending at a node in partition
$j$.  Clearly the sum of any row (or column) of ${\mathbf e}$, namely $a_i
= \sum_j e_{ij}$, corresponds to the fraction of links connected to
$i$.
For a random network, that does not exhibits any community structure, 
the expected value of the fraction of links within
partitions would be simply the probability that a link
begins at a node in $i$, $a_i$, multiplied by the fraction of links
that end at a node in $i$, $a_i$. So the expected number of
intra-community links is just $a_ia_i$. On the other hand, 
we know that the {\it real} fraction of links exclusively within a partition is
$e_{ii}$. So, we can compare the two directly and sum over all the
partitions in the graph, thus obtaining exactly the definition of modularity:

\begin{equation}
Q\equiv\sum_i(e_{ii} - a_i^2) ~~~.
\label{defsq}
\end{equation}

It is easy to see that if we take the whole 
network  as a single community, or if the network is a random one, we get the minimum value $Q=0$; 
on the other hand,  values approaching the maximum value $Q=1$ indicate strong community
structure. In practice, however, $Q$ never reaches the value $1$ and,
for networks with an appreciable subdivision in classes, it
usually falls in the range [$0.2$,$0.7$].  
\\
If we apply such a method to the Tepoztl\'an network, we find that the best subdivision
is the one in six groups $G_1$, $G_2$, $G_3$, $G_4$, $G_5$ and $G_6$ 
shown in Fig.2. This division has the considerable modularity value of $Q=0.534$. 
We report the detailed list of the members of the six groups in Table 1.
It is noticeable that such a subdivision reproduces quite well (even if not exactly) 
the real scientific groups of the
Tepoztl\'an collaboration network, at least according to our perception. Thus in the following
we will adopt it in order to characterize clusters of agents sharing the
same opinion in the context of the OCR model.
Now let us    integrate model  (\ref{OCR_eq1}) numerically over the TN. 
\\
In this case the $x(t)$'s become the opinions of the $N=49$ agents of the TN at time $t$ and 
the $\omega$'s represent their natural \textit{ocr}.
Of course it would be difficult to hazard any hypotesis about the conservative or progressist
natural inclinations of the TN people, thus we will randomly choose the $\omega$'s
from a uniform distribution $\omega_i \in [-0.5,0.5]$.
Finally, as already pointed out before, ${\bf K}_i$ and $k_i$ represent, respectively, the neighborhood and the
degree of node $i$-th.
\\
First of all, starting from a uniform distribution on opinions into $[-1,1]$, we have verified that 
also in this case a transition from an homogeneous phase toward a synchronized one occurs around
the critical value $\sigma_c \sim 1.4$ of the strength of interaction.
Such a behavior is visible in Fig.1, where we plot with full circles the order parameter $R_\infty$ versus $\sigma$
for the OCR model on the Tepoztl\'an network, averaged over 100 realizations and 
compared with that one obtained for the fully connected network
(full line) discussed before. Notice that, in the TN case, the transition is smoother and the curve 
fluctuates more than the other one. 
This is due to the small number of nodes of the collaboration network, N=49,
if compared to that one of the fully connected system, N=1000.
In the next paragraph we will see that a narrow window $1.5 < \sigma < 2.5$ inside 
the partially synchronized phase results to be the most interesting region
of the diagram. 
\begin{figure}
\begin{center}
\epsfig{figure=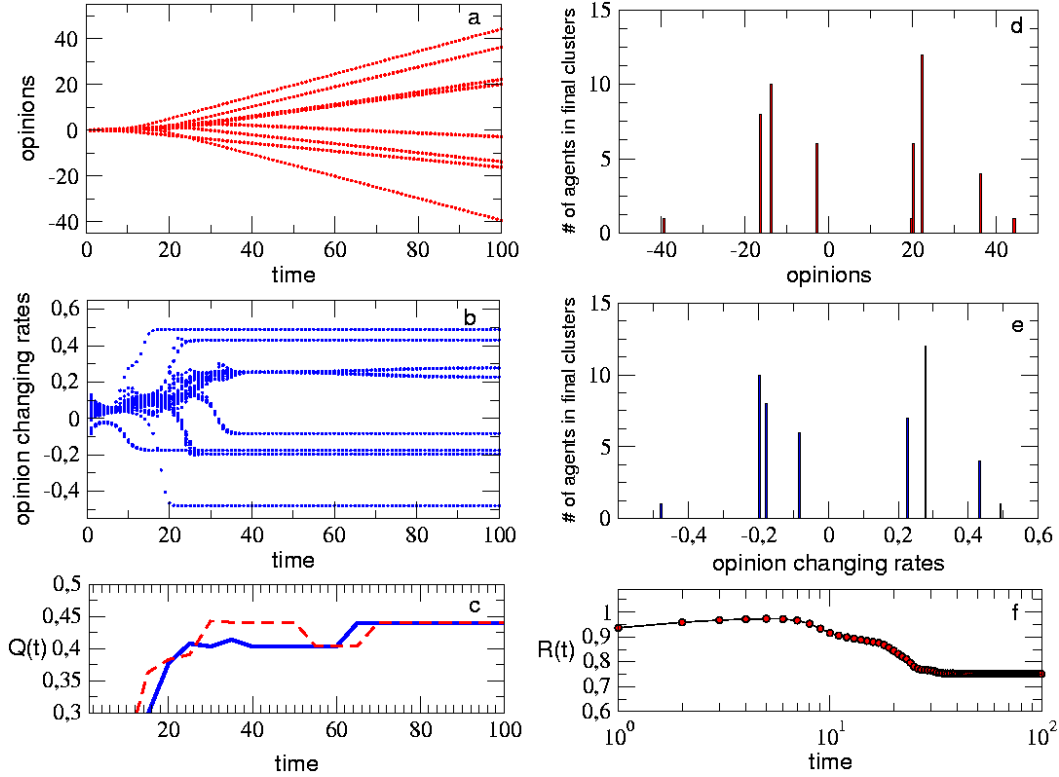,width=14truecm,angle=0}
\end{center}
\caption{Dynamical time evolution of groups synchronization in the OCR model on the Tepoztl\'an Network 
for $\sigma=2.0$ (one realization).
Starting from a metastable synchronized initial condition in the opinion space, panel (a), groups are spontaneously 
formed with components similar to that ones of the previous figure. In panel (b) the corresponding ocr time evolution is plotted. The modularity $Q(t)$ defined in eq. (2) is plotted vs time in panel (c) for the opinions (dashed line) and ocr (full line). In (d) and (e) the asymptotic  cluster distribution of panels (a) and (b) are plotted respectively. Finally in panel (f) the time evolution of the order parameter $R$ is plotted. See text for furtther details. 
}
\end{figure}

\subsection{Evolution of an initial state of synchronized opinions}

In Fig.3 we show the time evolution of the Tepoztl\'an network community for
$\sigma=2.0$ and for an initial state with all the members sharing the same opinion $x_i(0)=0 ~\forall ~i$.
The natural \textit{ocr} have been randomly chosen again in the range
$\omega_i \in [-0.5,0.5]$. In panels (a) and (b) we plot
the evolution of the opinions and of the opinion changing rates over 100 time steps 
(please notice that the uniformly distributed natural \textit{ocr} $\omega_i$ 
represent also the initial conditions for the \textit{ocr} variables $\dot{x}_i(t)$).
We can see that, after an initial short metastable transient in which both the opinions and
the opinions changing rates stay synchronized (in fact $R(t) \sim 1$, as shown in panel (f)),
the system rapidly clusterizes with a branching process strongly affected by the topology
of the network. The respective asymptotic stationary configurations of clusters are shown in panels (d) and (e).
Finally, in panel (c) the step-by-step modularities $Q$ for both the opinion clusters
configurations (dashed line) and the \textit{ocr} clusters configurations (full line) are also reported.
From the value $Q=0$ that characterizes the metastable configuration with only one large cluster,
both  modularities increase to a value $Q \sim 0.44$ in the asymptotic stationary state (d)-(e).
It is important to notice that these two final configurations, since they are time invariant, 
must necessarily  consist of the same number of clusters (even if arranged in a different order) 
with the same people inside, otherwise they would be dynamically unstable. 
\begin{table}
\begin{center}

\caption{ Asymptotic clusters configuration shown in panel (e) of Fig.3}
\begin{tabular} {ll}
\hline 
\hline
Group 1 (0.49)   & Sciortino(0.49)\\
Group 2 (0.43)   & Abe(0.49), Campa(0.42), Rajagopal(0.38), Tirnakli(0.28)\\
Group 3 (0.28)   & Andronico(-0.28), Baldovin(0.34), Caruso(-0.20), Crucitti(0.21),\\
                 & Latora(0.40), Pluchino(0.32), Rapisarda(0.34), Tadic(0.41),\\
		 & Beck(0.07), Cohen(0.40), Rizzo(0.08), Swinney(0.42)\\
Group 4 (0.23)   & Bukman(0.17), Indekeu(0.36), Widom(0.47), Stella(0.19),\\
                 & Dawson(-0.07), A.R.Plastino(0.15), Mukamel(0.26)\\
Group 5 (-0.08)  & Barr\'e(-0.07), Bouchet(0.01), Dauxois(0.28), Giansanti(-0.12),\\
                 & Ruffo(-0.16), Yamaguchi(0.12)\\
Group 6 (-0.17)  & J.Andrade(-0.46), Coniglio(-0.11), Franzese(-0.32), Havlin(-0.04),\\
                 & Herrmann(-0.21), Leyvraz(-0.37), Mantegna(0.21), Stanley(-0.21)\\
Group 7 (-0.20)  & Anteneodo(-0.20), Borland(-0.24), Gell-Mann(-0.23),\\         
                 &Tamarit(-0.40), Thurner(-0.30), Tsallis(-0.19),\\ &Hernandez-Saldana(-0.28),Majoral(-0.27),\\
		 &Moyano(-0.03), Robledo(-0.17)\\
Group 8 (-0.48)  & A.Plastino(-0.48)\\

\hline
\end{tabular} 
\end{center}
\end{table} 

The interesting result of the simulation is that the clusters of panel (d)-(e) show
a good overlap with the real communities of the Tepoztl\'an network shown in Fig.2. 
This can be verified comparing the groups' structure of Table 1 with that one 
reported in Table 2, representing the 
final clusters configuration of the \textit{ocr} plotted in panel (e).
In correspondence of each group in Table 2 the \textit{ocr} value of the cluster is reported,
together with the \textit{ocr} of each member of the cluster itself
(notice that the \textit{ocr} of each cluster is rather close to the average 
of the $\omega$'s of its members).
It is important to stress that people
in each final cluster, evidently due to the strong influence of the network topology on the
mutual interactions, manage to maintain synchronized opinions 
\textit{despite} their different natural inclinations $\omega_i$ (see panel (a)). 
On the other hand, during the branching evolution people with similar $\omega_i$ 
tend to merge together in the same cluster while people with very different $\omega_i$ 
tend to escape from a given cluster (see panel (b)). 
The competition between these two opposite effects can  explain why some of the groups  well distinct  in Table 1, merge in Table 2 (e.g. G1-G2 or G3-G5), 
why new groups come up (see Group 2 in Table 2) 
and why agents with a very high absolute value of $\omega$  
run alone at the extremes of the \textit{ocr} range. 
Consequently the resulting modularity of the asymptotic clusters configuration shown in Table 2 ($Q \sim 0.44$)
is smaller than the modularity of the "real" configuration ($Q=0.53$).
\\
We checked that even if the details of the simulation of Fig.3 change for different runs 
with different random realizations of $\omega_i \in [-0.5,0.5]$, and the final modularity can
vary too, the global picture described before remains qualitatively the same,
provided that the interaction strength would stay around $\sigma=2.0$.
For values noticeably higher or lower than $2.0$ the system, respectively, 
either remains synchronized forever or quickly esplodes in very small clusters.

\begin{figure}
\begin{center}
\epsfig{figure=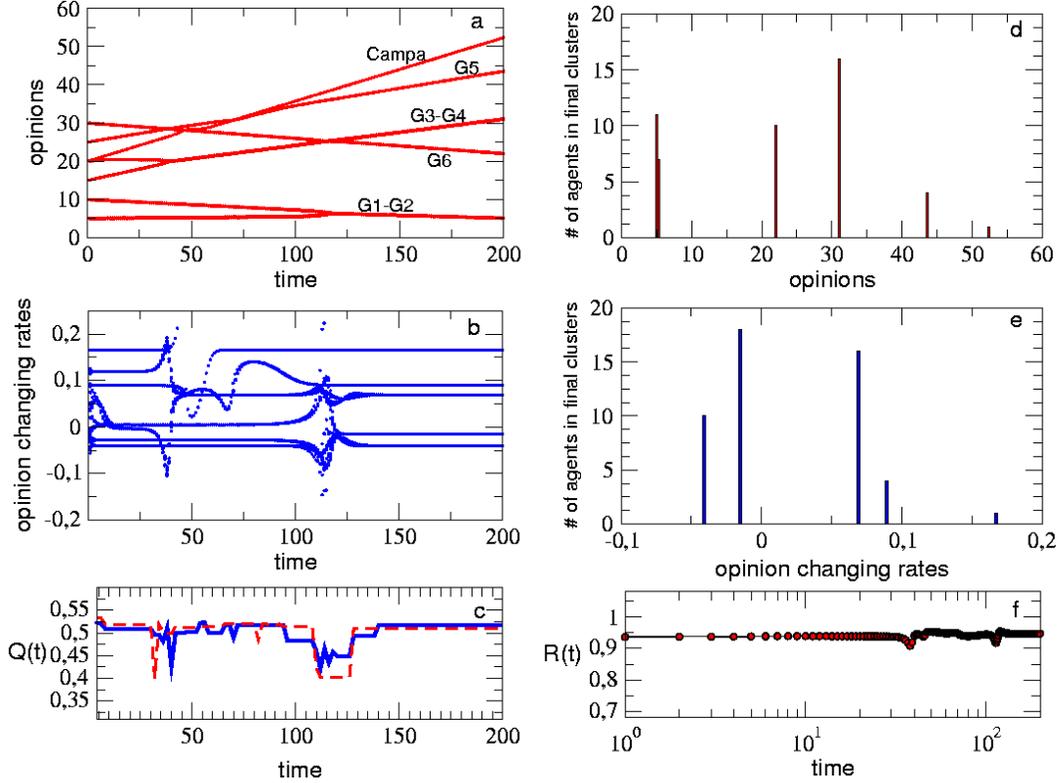 ,width=14truecm,angle=0}
\end{center}
\caption{Time evolution of the OCR model on the Tepoztlàn collaboration network.
In the opinion space (a) the agents start into
six synchronized groups corresponding to the six communities
of Table 1, from $G1$ - at the bottom with x(0)=5 - to $G6$ - at the top with x(0)=30 - (see also Table 3).
On the other hand, the agents' natural rates (i.e. the initial conditions in panel (b)) are randomly chosen.  
See caption of the previous figure and text for further details. 
}
\end{figure}

\subsection{Evolution of the coauthorship groups in the opinion space}

In this subsection we want to explore what happens if we start the system 
with six different clusters of synchronized opinions corresponding to the six real communities 
of the Tepoztl\'an network, in order to see how the OCR dynamics affects the stability of the groups.
Of course also in this case the details of the simulations will depend on the 
initial distribution of the $\omega$'s. 
But, again, we are mainly interested in studying the global picture emerging from 
the dynamical competition between the different natural 
\textit{ocr} of the agents in each group and the opinion constraints imposed by the topology
of the network.
\\
In the simulation shown in Fig.4 we set again $\sigma=2$. 
Then we chose, for each one of the six groups of Table 1, a different initial opinion, 
common for all its members, while the natural rates $\omega$'s were randomly selected in the range $[-0.2,0.2]$. 
The initial opinion of each group is reported in Table 3.  
Like in the previous section, in panels (a) and (b) of Fig.4 we  show the time evolution of the
opinions $x_i(t)$ and the opinion changing rates $\dot{x}_i(t)$, while in panel (c) the respective modularities
are plotted. Finally, in panels (d) and (e), we  show the asymptotic clusters configuration for
both the opinions and the \textit{ocr}, and in panel (f) the time behavior of the order
parameter $R(t)$ is reported.
\begin{table}
\begin{center}
\caption{Initial position of the six real groups in the opinion space, see
panel (a) of  Fig.4}
\begin{tabular} {ll}
\hline 
\hline
G1: x(0)=5  -  & Bukman, Hernandez-Saldana, Indekeu, Majoral, Moyano,\\
               & Robledo, Widom\\
G2: x(0)=10 -  & Abe, Anteneodo, Borland, Gell-Mann,A.Plastino, A.R.Plastino,\\
               & Rajagopal, Tamarit, Thurner, Tirnakli, Tsallis\\
G3: x(0)=15 -  & Andronico, Baldovin, Caruso, Crucitti, Latora, Pluchino,\\
               & Rapisarda, Tadic\\
G4: x(0)=20 -  & Barr\'e, Bouchet, Dauxois, Campa, Giansanti, Mukamel, Ruffo,\\
               & Stella, Yamaguchi\\
G5: x(0)=25 -  & Cohen, Beck, Rizzo, Swinney\\
G6: x(0)=30 -  & J.Andrade, Coniglio, Dawson, Franzese, Havlin,\\
               & Herrmann, Leyvraz, Mantegna, Sciortino, Stanley\\
\hline
\end{tabular} 
\end{center}
\end{table} 


\begin{table}
\begin{center}
\caption{Final positions and configurations of the clusters in panel (d) of Fig.4}
\begin{tabular} {ll}
\hline 
\hline
Group 1:            & Bukman(0.07), Hernandez-Saldana(-0.11), Indekeu(0.15),\\
x(200)=5.05         & Majoral(-0.10), Moyano(-0.01), Robledo(-0.07), Widom(0.19),\\
		    & Abe(0.20), Anteneodo(-0.08), Borland(-0.09), Gell-Mann(-0.09),\\
		    & A.Plastino(-0.19), A.R.Plastino(0.06), Rajagopal(0.15),\\
		    & Tamarit(-0.16), Thurner(-0.12), Tirnakli(0.11), Tsallis(-0.07)\\ 
Group 2:            & J.Andrade(-0.18), Coniglio(-0.05), Dawson(-0.03),Franzese(-0.13),\\  
x(200)=21.95        & Havlin(-0.01), Herrmann(-0.08), Leyvraz(-0.14), Mantegna(0.08),\\ 
	            & Sciortino(0.20), Stanley(-0.08)\\					      
Group 3:            & Andronico(-0.11), Baldovin(0.13), Caruso(-0.08), Crucitti(0.08),\\
     x(200)=31.02   & Latora(0.16), Pluchino(0.12), Rapisarda(0.14), Tadic(0.17)\\
                    & Barr\'e(-0.02), Bouchet(0.00), Dauxois(0.11), Giansanti(-0.04),\\
                    & Mukamel(0.10), Ruffo(-0.06), Stella (0.07), Yamaguchi(0.05)\\
Group 4:            & Cohen(0.16), Beck(0.02), Rizzo(0.03), Swinney(0.17)\\
x(200)=43.54  \\
Group 5:   & Campa(0.16)\\ 
x(200)=52.33 \\
\hline
\end{tabular} 
\end{center}
\end{table} 


Let us  follow  the time evolution of the opinion clusters in panel (a) of Fig.4 along 200 time steps.
One can see that groups G1 and G2 merge at $t\sim120$, due to their initial position and 
to their similar average \textit{ocr}. Almost immediately a fast agent (Campa,  with $\omega=0.16$)
leaves the group G4 and  goes ahead to its natural \textit{ocr}, 
resisting alone until the end of the simulation (see next Table 4).
Meanwhile, what remains of G4 merges with G3 at $t=38$, 
and at $t=120$ the new resulting group survives to the collision with group G6,
 already survived to previous superpositions with G5 and Campa around $t=50$.
On the other hand, looking to panel (b), the opinion changing rates 
follow a more pronounced branching evolution, where the agents rearrange their   
\textit{ocr} until a final stationary state is reached.
However, as previously stressed, the asymptotic configurations shown in panels (d) and (e) 
must be the same.
Notice that the two main rearrangements in panel (b), around $t\sim50$ and $t\sim120$,
correspond to the main cluster collisions in panel (a) and also to the sudden falls
of modularity in panel (c).
\\
In Table 4 we report the detailed composition of the
asymptotic opinion clusters of panel (d), together with their position in the opinion space.
Comparing it with the starting configuration of Table 3, we see that 
the real groups of the Tepoztl\'an network are quite stable, at least for $\sigma=2$,
despite the random choice of the natural changing rates of their members.
Apart from one agent of group G4, and apart from a couple of fusions (G1-G2 and G3-G4), 
the structure of the groups seems to have been preserved by the dynamics,
surviving to various collisions in the opinion space.
This is confirmed by the final modularity $Q=0.51$, not much smaller then
the original one $Q=0.53$, and also by the constant behavior of the 
order parameter in panel (f).
Changing the initial  distribution of  \textit{ocr},  the evolution of the each member can change, but the qualitative behaviour  is the same.

\section{Conclusions}
In this paper we have discussed opinion dynamics in a real social network, by 
considering the OCR model introduced in ref.\cite{ocr}. In particular we have investigated a 
network of scientific coauthorship, inspired to Alberto Robledo's collaborations.  
The results demonstrate that the topology of the network is a fundamental ingredient 
in the opinion dynamics and in the evolution of the composition of scientific groups. 
The use of the OCR model seems very promising for studying the dynamics of opinion formation.  
Further analysis with different kinds of networks are in order to draw  more definitive conclusions.   

\section{Acknowledgements}

We  would like to dedicate this paper to  Alberto Robledo, wishing him  a long 
and  very productive  continuation of his brilliant  academic career.

Two of us  (A.P.) and (A.R.) would like to thank the organizers for the 
warm hospitality and the 
financial support.



\begin{thebibliography}{00}


\bibitem{ocr}
A.Pluchino, V.Latora and A.Rapisarda
{\it Int. J. Mod. Phys. C}, {\bf 16}, No.4, 515-531 (2005) 

\bibitem{general} 
L.D.Kiel, E.Elliot {\it Chaos theory in the social sciences} The University of Michigan Press  (1996) 
L.Tesfatsion, {\it Economic agents and markets as emergent phenomena}, PNAS, vol.99 (2002) 7191-7192
J.M.Epstein and R.Axtell, {\it Growing Artificial Societies} (1996) MIT Press, Cambridge MA
W.B.Arthur, S.N.Durlauf and D.A.Lane, {\it The Economy as an Evolving Complex System II}, 
Addison-Wesley, Reading, MA, vol. XXVII 
R. Axelrod, {\it J. Conflict Resolut.} {\bf 41}, 203 (1997).
S. Fortunato, {\it On the Consensus Threshold
for the Opinion Dynamics of Krause-Hegselmann}, cond-mat/0408648
at www.arXiv.org, to appear in {\it Int. J. Mod. Phys. C} {\bf 16}, issue 2 (2005).

\bibitem{nature-soft} 
{\it In praise of soft science} Editorial (2005), Nature, 435(7045), 1003.

\bibitem{rev1} R.\ Albert and  A.-L.\ Barab\'{a}si, Rev. Mod. Phys. \textbf{74}, 47 (2002).

\bibitem{rev2} M.E.J. Newman, SIAM Review {\bf 45}, 167 (2003).

\bibitem{rev3} S. Boccaletti, V. Latora, Y. Moreno, M. Chavez and D.-U. Hwang, Phys. Rep. {\bf 424}, 175 (2006)
and refs. therein.

\bibitem{stauff1}
D.Stauffer, {\it Sociophysics Simulations} (2003) in 'Computer Simulations', Co-published by the IEEE CS and the AIP
D.Stauffer, {\it Sociophysics Simulations II: Opinion Dynamics} (2005) arXiv:physics/0503115

\bibitem{HK} 
R. Hegselmann and U. Krause, {\it Journal of Artificial Societies and
Social Simulation} {\bf 5}, issue 3, paper 2 (jasss.soc.surrey.ac.uk) (2002).

\bibitem{SDG} 
K. Sznajd-Weron and J. Sznajd, {\it Int. J. Mod. Phys. C} {\bf 11}, 1157 (2000).
G. Deffuant, D. Neau, F. Amblard and G. Weisbuch, {\it Adv. Complex Syst.} {\bf 3}, 87 (2000).
S. Galam, {\it Physica} A {\bf 336} (2004) 49 and refs. therein.

\bibitem{bennaim} E. Ben-Naim, P. Krapivsky and S. Redner, {\it Physica D} {\bf 183},
190 (2003).

\bibitem{stauff2}
A.T.Bernardes, D.Stauffer and J.Kertész, {\it Eur.Phys.J.} B {\bf 25}, 123-127 (2002)
M.C.Gonzales,A.O.Sousa and H.J.Herrmann, {\it Int. J. Mod. Phys. C}, {\bf 15}, No.1, 1-13 (2004)
D.Stauffer, {\it Montecarlo simulations of the Sznajd model}, 
{\it J.of Artificial Societies and Social Simulations} (2002) vol.5, no.1 

\bibitem{vector}
S.Fortunato, V.Latora, A.Pluchino and A.Rapisarda
 International Journal of Modern Physics C 16, 1535 (2005)
and  A. Pluchino,  V.Latora, A. Rapisarda, Eur. Phys. Journ.  B {\bf 50} (2006) 169.



\bibitem{boc02}  S. Boccaletti, J. Kurths, D.L. Valladares, G. Osipov and C.S. Zhou, Phys. Rep. {\bf 366}, 1 (2002).

\bibitem{kuramoto_model}
Y. Kuramoto, in {it International Symposium on Mathematical Problems in
Theoretical Physics}, Vol.~39 of {\it Lecture Notes in Physics}, edited 
by H. Araki (Springer-Verlag, Berlin, 1975).

\bibitem{k1}  Y. Kuramoto, {\it Chemical Oscillations, Waves, and Turbulence} (Springer, Berlin, 1984).

\bibitem{strogatz}
S.~H Strogatz, {\it Physica D}, {\bf 143} 1 (2000). 

\bibitem{GN} M.~Girvan and  M.~E.~J. Newman,  Proc. Natl. Acad. Sci. USA \textbf{99}, 7821  (2002). 
M.E.J. Newman and M. Girvan {\em Phys. Rev. E}, {\bf 69}, 026113 (2004).



\end{thebibliography}
\end{document}